\documentclass[preprint,12pt]{elsarticle}
\usepackage{amssymb}
\usepackage{amsmath}
\usepackage{amsthm}
\usepackage{amsfonts}
\theoremstyle{plain}

\newcommand{\myvec}[1]{\vec{\vphantom{A}#1}}
\usepackage{setspace}
\usepackage{algorithm}
\usepackage{algpseudocode}
\usepackage{algorithmicx}

\usepackage[colorlinks, linkcolor=red, anchorcolor=green, citecolor=blue]{hyperref}
\usepackage{url}
\usepackage{graphicx}
\graphicspath{{figures/}}
\usepackage{subfigure}
\usepackage{float}
\usepackage{enumerate}
\usepackage{enumitem}
\newlist{steps}{enumerate}{1}
\setlist[steps, 1]{label = Step \arabic*.}
\usepackage{diagbox}
\usepackage{tabularx}
\usepackage{tabu}
\usepackage{colortbl}
\definecolor{mygray}{gray}{.9}
\definecolor{mypink}{rgb}{.99,.91,.95}
\definecolor{mycyan}{cmyk}{.3,0,0,0}
\usepackage{tikz}
\usetikzlibrary{shapes.geometric, arrows}
\tikzstyle{startstop} = [rectangle, rounded corners, minimum width = 2cm, minimum height=1cm,text centered, draw = black, fill = red!30]
\tikzstyle{io} = [trapezium, trapezium left angle=70, trapezium right angle=110, minimum width=2cm, minimum height=1cm, text centered, draw=black, fill = blue!30]
\tikzstyle{process} = [rectangle, minimum width=3cm, minimum height=1cm, text centered, draw=black, fill = orange!30]
\tikzstyle{decision} = [diamond, aspect = 3, text centered, draw=black, fill = yellow!30]
\tikzstyle{arrow} = [->,>=stealth]
\tikzstyle{line} = [thick, -, >= stealth]
\usepackage{bm}
\usepackage{geometry}
\journal{Results in Applied Mathematics}
\begin{document}

\begin{frontmatter}

\title{A hybrid recommendation algorithm based on weighted stochastic block model\tnoteref{mytitlenote}}
\tnotetext[mytitlenote]{This work is supported by Guangdong Province Student’s Platform for Innovation and Entrepreneurship Training Program (Grant No.1055914152), and The Challenge Cup of the Ministry of Education of China (Grant No.15112005).}

\author[label1]{Yuchen Xiao\corref{corr}}
\author[label2]{Ruzhe Zhong}
\address[label1]{Department of Mathematics, Jinan University, Guangzhou, China}
\address[label2]{Department of Mathematics, The Hong Kong University of Science and Technology, Hong Kong, China}
\cortext[corr]{Corresponding author, E-mail address: fkxych@stu.jnu.edu.cn}

\begin{abstract}
Hybrid recommendation usually combines collaborative filtering with content-based filtering to exploit merits of both techniques. It is widely accepted that hybrid filtering outperforms the single algorithm, thus it has been the new trend in electronic commerce these years. In this paper, we propose a novel hybrid recommendation system based on weighted stochastic block model (WSBM). Our algorithm not only makes full use of content-based and collaborative filtering recommendation to solve the cold-start problem but also improves the accuracy of recommendation by selecting the nearest neighbor with WSBM. The experiment result shows that our proposed approach has better prediction and classification accuracy than traditional hybrid recommendation. 
\end{abstract}


\end{frontmatter}
\section{Introduction}
With the development of information technology and network technology, the scale of information from the Internet has rapidly increased in recent years. How to filter overloaded information effectively and recommend useful information to users has become a hot issue in recommendation systems. The current recommendation systems algorithms contain content-based, social-based \cite{wang2014online}, context-aware, collaborative filtering \cite{candillier2007comparing,bobadilla2012collaborative}, knowledge-based \cite{shi2004intelligent,bobadilla2013recommender}, graph-based \cite{wei2013distinguishing} and hybrid recommendation \cite{porcel2012hybrid}. Among all the algorithms, content-based, collaborative filtering and hybrid recommendation are widely used.

Content-based recommendation system \cite{van2000using,salter2006cinemascreen} determined the preferences of users by their choices made before. Content-based recommendation extracts the description documents of items and determines whether the items suit users by comparing them with user's preference.

Collaborative filtering algorithm \cite{candillier2007comparing,su2009survey} is one of the most widely used and mature recommendation methods \cite{burke2002hybrid}. Its main idea is to use the historical rating data of the user's nearest neighbor to predict the rating of him or her, and then recommend the Top-N items for him or her. Breese divided the collaborative filtering algorithm into two parts: memory-based and model-based \cite{breese1998empirical}. Memory-based algorithm is simple but its accuracy will reduce with the increase of users, while model-based algorithm need to create new models frequently to adapt to the change of users or projects.
	
It is evident that different recommendation methods have different advantages and disadvantages, so hybrid recommendation \cite{xu2005content,borras2014intelligent} has become a commonly used algorithm in recommendation system. For example, Ariyoshi and Kamahara proposed a hybrid information recommendation method by applying singular value decomposition (SVD) on both collaborative filtering and content-based filtering respectively to reduce the cost of computation \cite{ariyoshi2010hybrid}. Lucas et al. divided the users into groups using personal demographic data (Demographic-based), content information of the items previously selected by the user (Content-based) and the information of other users (Collaborative filtering) \cite{lucas2013hybrid}. Rathachai et al. proposed a prediction model which is a combination of three scoring functions, and took collaborative filtering, community structure, and biological classification into account \cite{chawuthai2014link}. It can not only take full advantages of a variety of recommendation technologies but also avoid their disadvantages. The accuracy of the recommendation can be increased by using it.
	
According to the users' basic information and historical users' rating data, we propose a hybrid recommendation based on WSBM. Our hybrid recommendation contains two parts: the content-based part and the collaborative filtering part. The content-based part predicts the users' ratings through the similarities of item documentations, so as to recommend the new item. The collaborative filtering part combines the similarity of users' basic information with the similarity of historical users' rating data and constructed the overall similarity, which can achieve recommendation for new users. Moreover, in this part, we generate a user-user social network based on the same purchase records and adopt the WSBM to find the nearest neighbor, which improves the accuracy of the algorithm.

\section{Hybrid recommendation based on WSBM}
\subsection{Content-based rating prediction}
\subsubsection{Building the feature document of items}
Vector space model (VSM) is used in this paper to build the feature document of items. We firstly extract the description documents of items from the web, and remove the stop words and function words. According to tf-idf \cite{kim2014noise}, the feature weight of the description documents can be calculated, and the feature words above the given threshold $\sigma$ can be chosen to build a feature space. In the feature space, treat the feature words as key code and the weights as the values for the key code. Thus we get the document-word frequency matrix $X_{r\times s}$ (where $r$ is the number of feature words and $s$ is the number of the items contained in item set $N$). We use latent semantic analysis \cite{zhong2010novel} to map the document into a lower dimensional latent semantic space and we obtain an approximate matrix $X_M$ as follows
\begin{align}
\label{eq1}
X_M={U_M}{S_M}{V_M}^T.
\end{align}
\subsubsection{Predicting rating}
According to the approximate document-word frequency matrix $X_M$, we calculate the similarity between item $I_j$ and item $I_k$ in (\ref{eq2}) to obtain the similarity matrix for $j,k=1,2,\ldots,s$. To predict the rating of user $U_a$ for item $I_j$, we select the items which have been rated by user $U_a$ to form a reference set $I_c^*$. According to the similarity matrix and the reference set, we can predict the rating in (\ref{eq3})\cite{ghauth2010learning}
\begin{align}
\label{eq2}
sim(I_j,I_k)=\cos(I_j,I_k)=\frac{w_k\cdot w_j}{\lVert w_k\rVert\lVert w_j\rVert},\\
\label{eq3}
p_{a,j} = \frac{\sum\limits_{k\ne j,I_k\in I_c^*}sim(I_j,I_k)p_{a,k}}{\sum\limits_{k\ne j,I_k\in I_c^*}sim(I_j,I_k)}.
\end{align}
where $w_j$ is the $j^{th}$ column of $X_M$, $p_{a,j}$ is the rate of user $U_a$ for item $I_j$.
\subsection{Rating prediction of collaborative filtering}
\subsubsection{Selecting the nearest neighbor based on WSBM}
There are two commonly used methods to select the nearest neighbors: one is to select neighbors whose similarity to the target user is larger than a threshold \cite{kim2007effective}, the other is to search for the neighbors who have the greatest $N$ similarities to the target user \cite{bobadilla2011framework}. Both of the two methods are not universal on different datasets for the reason that the best recommendation can be achieved on different threshold or $N$.

As the community is considered to have a high correlation among users in social networks, we consider that users may have similar interests if they purchase same things. So we define the users as the vertexes of the network and if users have the same purchase, an edge is established between them, thus the numbers of same purchases are defined as the weights of edges. After constructing a weighted network, we adopt the WSBM to detect the community structure and to find the vertex’s nearest neighbors in the community. Without relying on the similarities, this method is more universal on different datasets.

Based on Stochastic Block Model, WSBM \cite{aicher2013adapting} defines the distribution of edge as two parts: the existence distribution and the weight distribution. For vertexes in the same community, to avoid heavy-tailed degree distributions, the model revises the probability of edge’s connection between one vertex to another according to the vertex degree.

In the WSBM, we define $A$ as the adjacency matrix of the weighted network $N$, and $A_{ij}$ as the weight of the edge between the vertexes $i$ and $j$. The integer $K$ denotes a fixed number of latent communities, and the vector $Z_{n\times 1}$ contains the community labels of each vertex. The WSBM defines a "bundle" of edges that run between each pair of communities ($kk'$) and assigns an edge existence parameter to each edge bundle $kk'$, which we represent collectively by the matrix $\theta_{K\times K}$. The existence probability of an edge $A_{ij}$ is given by the parameter $\theta_{Z_i}\theta_{Z_j}$ that depends only on the community memberships of vertexes $i$ and $j$. Therefore, the model is fully given by $\theta_{K\times K}$ and $Z_{n\times 1}$ \cite{aicher2014learning}.

WSBM models an edge's existence as a Bernoulli or binary random variable and an edge's weight using an exponential family distribution. With the Bernoulli distribution to simulate the existence distribution, vertex degree information is added into the "edge-propensity" parameter $\phi_i$ to each vertex. As a result, the existence probability of an edge $A_{ij}$ is a Poisson random variable with mean $\phi_i\phi_j\theta_{Z_i}\theta_{Z_j}$. Because the maximum likelihood estimate of each propensity parameter $\phi_i$ is simply the vertex degree $d_i$, by fixing $\phi_i=d_i$, we can obtain the likelihood function for this model
\begin{align}
P(A|Z,\theta)\propto\prod_{ij}\exp(A_{ij}\cdot\log\theta_{Z_iZ_j}-d_id_j\cdot\theta_{Z_iZ_j}),
\end{align}
which can be rewritten as
\begin{align}
P(A|Z,\theta)\propto\sum_{ij}T_{\epsilon}(A_{ij})\cdot\eta_{\epsilon}(\theta_{Z_iZ_j}).
\end{align}
where $T_{\epsilon}=(A_{ij},-d_id_j)$ is the sufficient statistics and $\eta_{\epsilon}=(\log\theta_{Z_iZ_j},\theta_{Z_iZ_j})^\top$ is the natural parameters.

With the exponential family distribution to simulate the existence of the distribution of the edges' weight, for the given parameters $Z$ and $\theta$, we assume that $A_{ij}$ is conditionally independent, thus $A_{ij}$ is an exponential random variable parametrized by $\theta_{Z_iZ_j}$ over domain $\mathcal X$. Hence the likelihood has the form of an exponential family
\begin{align}
P(A|Z,\theta)\propto\exp(\sum_{ij}T_{\omega}(A_{ij})\cdot\eta_{\omega}(\theta_{Z_iZ_j})).
\end{align}

Choosing an appropriate $(T_{\omega},\eta_{\omega})$, we can specify a stochastic block model, which weights are drawn from an exponential family distribution.

Then we may combine their contributions in the likelihood function via a simple tuning parameter $\alpha\in[0,1]$ that determines their relative importance in inference
\begin{align}
\log P(A|Z,\theta)=\alpha\sum_{ij\in E}T_{\epsilon}(A_{ij})\cdot\eta_{\epsilon}(\theta_{Z_iZ_j})+(1-\alpha)\sum_{ij\in W}T_{\omega}(A_{ij})\cdot\eta_{\omega}(\theta_{Z_iZ_j}),
\end{align}
where $E$ is the set of observed interactions (including non-edges) and $W$ is the set of weighted edges ($W\subset E$).

In order to deeply understand the structure of the network, a reasonable number of latent block structure $K$ should be chosen. Moreover, we need to learn the parameters $Z$ and $\theta$ by exploiting maximum-likelihood function. Steps of the algorithm are as follows
\begin{steps}
\item Determine the number of block structures $K$. Adopt the Bayesian model in search of the $K$ that maximizes the posterior probability $P(m|A)$, which is
\begin{align}
p(m|A)=\frac{p(A|m)p(m)}{\sum\limits_{m\in M}p(m,A)},
\end{align}
where $m$ denotes a specific model, and $M$ denotes the model set with different $K$.
\item Learn parameters $Z$ and $\theta$
\begin{itemize}[leftmargin=0 pt]
    \item (a)
According to Variational Bayesian Model, the log-likelihood function is
\begin{align}
\log A=L(q)+KL(q\lVert p(\cdot|A)),
\end{align}
where $\log A$ is a constant, and
\begin{align}
L(q)&=\int\sum\limits_Z q(Z,\theta)\log\frac{p(A,Z,\theta)}{q(Z,\theta)}d\theta,\\
KL(q\lVert p(\cdot|A))&=-\sum\limits_Z\int q(Z,\theta)\log\frac{p(Z,\theta|A)}{q(Z,\theta)}d\theta.
\end{align}
    \item (b)
Maximize the lower bound $L(q)$, thus we get the minimum $KL$ distance, then $q(Z,\theta)$ is the nearest distribution to the posterior distribution $p(Z,\theta|A)$.
    \item (c)
Based on total variation theory
\begin{align}
q(Z,\theta)=q_\theta(\theta)\prod_{i=1}^n q_Z(Z_i),
\end{align}
and by using Expectation–maximization algorithm (EM algorithm) to optimize $q_Z(Z_i)$ within E-step and optimize $q_\theta(\theta)$ within M-step, both the optimal parameters $Z$ and $\theta$ can be figured out.
\end{itemize}
\end{steps}
Through the process above, all users in the network will be classified into $K$ communities. As for a certain user, the rest users in the same community are regarded as its nearest neighbors. We use $NN_a$ to represent the set of nearest neighbors of the user $U_a$.

With the increase of web users, it is necessary to update the network. We can define a threshold of new users. When the number of new users reaches a specified limit, redetect the communities. Owing to the offline operation of community detection, the efficiency of online recommendation will not be affected.
\subsubsection{Similarity calculating}
\begin{enumerate}[leftmargin=1 pt]
\item Similarity Based on Users’ Basic Information

Albert et al. find that the user's basic information such as gender, age, occupation, cultural background would have a relatively large impact on its interest preference \cite{albert2000error}. We calculate the similarity of basic information between target user $U_a$ and $U_i$ by constructing vector $\overrightarrow{use_i}$ from the basic information of user $U_i$ \cite{zhang2014collaborative},that is
\begin{align}
sim_1(U_i,U_a)=\cos(U_i,U_a)=\frac{\overrightarrow{use_i}\cdot\overrightarrow{use_a}}{\lVert\overrightarrow{use_i}\rVert\lVert\overrightarrow{use_a}\rVert}.
\end{align}
\item Similarity Based on Users’ Rating

We collect the rating, evaluation of user’s and then clean, convert and entry them. Eventually we form a data matrix, containing each item evaluated by users $A=(R_{ij})_{m\times n}$, where $R_{ij}$ stands for the existing rating of user $U_a$ for item $I_j$, the formula presents as follows
\begin{align}
sim_2(U_i,U_a)=\cos(\myvec{i},\myvec{a})=\frac{\myvec{i}\cdot\myvec{a}}{\lVert\myvec{i}\rVert\lVert\myvec{a}\rVert},
\end{align}
where $\myvec{i}$ and $\myvec{a}$ are the rating vectors of user $U_i$ and $U_a$ respectively.
\end{enumerate}

Now we can define the integrated similarity $sim(U_i,U_a)$ as
\begin{align}
\label{eq4}
sim(U_i,U_a)=
\begin{cases}
sim_1(U_i,U_a),&\text{$U_a$ is a new user,}\\
\alpha sim_1(U_i,U_a)+\beta sim_2(U_i,U_a),&\text{$U_a$ is an old user,}
\end{cases}
\end{align}
where $\alpha$ and $\beta$ are adjustable parameters which show the contribution of users’ basic information and users’ rating to the recommendation system.
\subsubsection{Predicting rating}
The reference set $U_a^*$ is formed by users included in $NN_a$ as well as having evaluated on $I_j$. We use all ratings of $U_i$ $(U_i\subset U_a^*)$ to predict $pp_{a,j}$ (the rating of user $U_a$ for item $I_j$) \cite{xie2013similarity}, which is
\begin{align}
pp_{a,j}=\overline{R_a}+\frac{\sum\limits_{U_i\in NN_a}sim(U_i,U_a)(R_{i,j}-\overline{R_i})}{\sum\limits_{U_i\in NN_a}\lvert sim(U_i,U_a)\rvert},
\end{align}
where $R_{i,j}$ is the rating of $U_i$ for item $I_j$, $\overline{R_i}$ and $\overline{R_a}$ are the average rating of $U_i$ and $U_a$ respectively.
\subsection{Producing recommendation}
Therefore, the final rating that target user $U_a$ given to item $I_j$ can be defined as follows
\begin{align}
\label{eq5}
P_{a,j}=\gamma_1 p_{a,j}+\gamma_2 pp_{a,j},
\end{align}
where $\gamma_1$ and $\gamma_2$ are adjustable parameters which shows the contribution of item-based and collaborative filtering to the recommendation system. In the initial stage of system operation, content-based is more accurate than collaborative filtering so $\gamma_1>\gamma_2$. And then collaborative filtering will be more accurate with the increasing number of users and rating, so $\gamma_1<\gamma_2$. Ultimately we recommend the items of which the ratings ranked top N to user $U_a$.
\subsection{Improving the rating of user}
To some extent, the ratings of users can describe the preferences of them and the possible range is $0$ to $5$. The higher the ratings are the more users are interested in the items. If the recommendation system only considers users’ ratings, there might be some simulated trading to improve the ratings which will reduce the accuracy of recommendation. In addition to this, the ratings mainly depend on the first impression so most of the information on quality cannot be reflected. If we only consider the ratings, it is difficult to distinguish the qualities of items thus the accuracy of recommendation will be inaccurate.

Additional comments can reflect the follow-up and long-term quality evaluation for items and they also synthesized feedbacks for service attitude and level. Therefore we can take the additional comments into account when calculate similarity, and the steps of additional comments processing are presented as follows
\begin{steps}
\item Extract the keywords of original comments scored ($1$ to $5$) to form the keyword set ($K_1$ to $K_5$), and then take the union set of them to form the overall keywords set $K^*$.
\item Calculate the weights of words in $K_l$ according to TF-IDF and construct the word frequency vector of rating $l$ which we denote as $V_l$ for $l=1,2,\ldots,5$.
\item Extract the keywords of additional comment of user $U_a$ for item $I_j$, then calculate the weights of these words, thus we get the frequency vector of the additional comment which we denote as $AV_j$ for $j=1,2,\ldots,s$.
\item Calculate $sim(AV_j,V_l)$ and denote the rating of the additional comment as the score $l^*$ when $sim(AV_j,V_l)$ is the highest, thus we get the appended rating of user $U_a$ for item $I_j(AR_{a,j})$.
\item Therefore, we get the final rating $FR_{a,j}$ as follows
\begin{align}
FR_{a,j}=\eta_1R_{a,j}+\eta_2AR_{a,j},
\end{align}
where $\eta_1$ and $\eta_2$ are adjustable parameters which show the contribution of users’ rating and users’ additional comments to the final ratings. For those without additional comments, they can be thought of having no particular like or dislike, so we define the appended rating of those people as a medium grade 3.
\end{steps}
\subsection{Algorithm design}
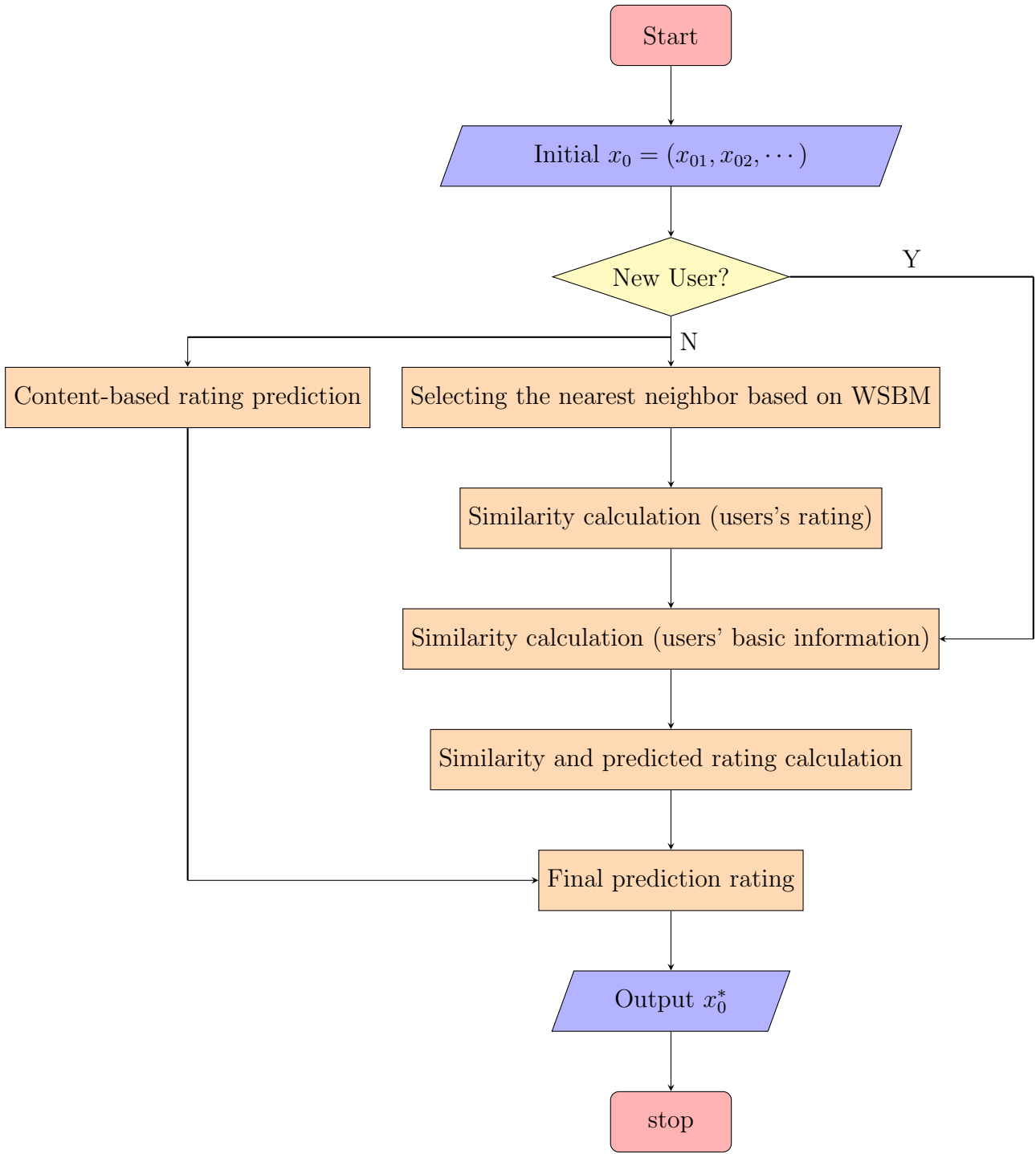
\begin{figure}[H] 
\centering 
\begin{tikzpicture}[node distance=2cm]
\node (start) [startstop]
{Start};
\node (in1) [io, below of = start]
{Initial $x_0=(x_{01},x_{02},\cdots)$};
\node (dec1) [decision, below of = in1]  
{New User?};
\node (pro2) [process, below of = dec1, xshift = -8cm]
{Content-based rating prediction};
\node (pro3) [process, below of = dec1]
{Selecting the nearest neighbor based on WSBM};
\node (pro4) [process, below of = pro3]
{Similarity calculation (users's rating)};
\node (pro5) [process, below of = pro4]
{Similarity calculation (users' basic information)};
\node (pro6) [process, below of = pro5]
{Similarity and predicted rating calculation};
\node (pro7) [process, below of = pro6]
{Final prediction rating};
\node (out1) [io, below of=pro7]
{Output $x_0^*$};
\node (stop) [startstop, below of=out1]
{stop};

\draw [arrow] (start) -- (in1);
\draw [arrow] (in1) -- (dec1);
\draw [line]  (dec1) --node [above] {Y} (6,-4);
\draw [line]  (6,-4) -- (6,-10);
\draw [arrow] (6,-10) -- (pro5);
\draw [line]  (0,-5) -- (-8,-5);
\draw [arrow] (-8,-5) -- (pro2);
\draw [line]  (pro2) -- (-8,-14);
\draw [arrow] (-8,-14) -- (pro7);
\draw [arrow] (dec1) --node [right] {N} (pro3);
\draw [arrow] (pro3) -- (pro4);
\draw [arrow] (pro4) -- (pro5);
\draw [arrow] (pro5) -- (pro6);
\draw [arrow] (pro6) -- (pro7);
\draw [arrow] (pro7) -- (out1);
\draw [arrow] (out1) -- (stop);
\end{tikzpicture}
\caption{Flow chart of hybrid recommendation algorithm}\label{fig.flowchart}
\end{figure}

Our hybrid recommendation system based on WSBM is a combination of content-based filtering algorithm and collaborative filtering algorithm. The algorithm is expressed as follows
\begin{steps}
\item Determine whether the user is a new user. If yes, go to \ref{step5} Otherwise, go to \ref{step2}
\item Content-based rating prediction. Calculate the similarity between items after extracting feature documents of items. Then predict the rating $p_{a,j}$ of $U_a$ for item $I_j$ by using historical rating data. \label{step2}
\item Selecting the nearest neighbor based on WSBM. Generate a user-user social network based on the same purchase records and structure the nearest neighbor set $NN_a$ of $U_a$ based on WSBM.
\item Calculate the similarity $sim_2(U_i,U_a)$ by using the historical rating of $U_a$ and $U_i$ $(U_i\in NN_a)$.
\item Calculate the similarity $sim_1(U_i,U_a)$ by using the basic information of $U_a$ and others. \label{step5}
\item Calculate the integrated similarity $sim(U_i,U_a)$ which is defined by (\ref{eq4}). Then predict the rating $pp_{a,j}$ of user $U_a$ for item $I_j$.
\item Calculate the final predicted rating $P_{a,j}$ which is defined by (\ref{eq5}).
\item Producing Recommendation. Recommend the items of which the ratings rank top N to user $U_a$.
\end{steps}
The whole process is shown in Figure \ref{fig.flowchart}.

\section{Experiments}
\subsection{An example of Improved Rating}
In order to prove the improvements of the ratings, we give a small example. We extract the 50000 ratings with comments from a well-known Chinese online shopping website: Suning (\url{http://www.suning.com/}) through the web crawler technology. We prove the correctness of our algorithm with 25 additional comments and their initial ratings, improved ratings and additional comments are shown in Table \ref{table1}.

\begin{table}[H]
\footnotesize
\centering
\caption{The evaluation of improving ratings}\label{table1}
\begin{tabular}{ccl}
\hline
 Initial ratings & Improved ratings & Appended Comments \\
\hline
\rowcolor{mygray}
5 & 3.5 & Seems like a pregnant! \\
5 & 4.5 & A pilling sweater! Poor quality! \\
\rowcolor{mygray}
4 & 4.5 & Not bad! I like it. \\
5 & 4.5 & It looks pretty good, but the color began to fading no longer than half a month! \\
\rowcolor{mygray}
5 & 4.5 & Warm but rough! \\
4 & 4.5 & An ideal piece! It is appropriate for autumn wear. \\
\rowcolor{mygray}
5 & 4.5 & It has thread breakage. \\
5 & 3 & Three stars. Look before you buy it! \\
\rowcolor{mygray}
5 & 4.5 & I love it, but it is more expensive than that in Tmall. \\
5 & 5 & Nice! I'll be back! \\
\rowcolor{mygray}
5 & 4 & Not so good. Little holes appeared at the sleeves. It might not be worth it. \\
5 & 4.5 & I placed the order two weeks ago but I haven't received it yet! What should I do? \\
\rowcolor{mygray}
5 & 4 & It is worn-out though I only wear a few times! \\
5 & 5 & Good quality! And it fits me very well! My friends think highly of it! \\
\rowcolor{mygray}
4 & 4.5 & So far so good. Not bad. \\
1 & 1 & They make a fool of the customers! Be careful of it! \\
\rowcolor{mygray}
5 & 4 & Uncomfortable! It has an unmatched waist line. \\
4 & 4.5 & A fair price! \\
\rowcolor{mygray}
2 & 3.5 & It is worth buying. \\
4 & 4.5 & Warm clothes! My parents are fond of them. We'll be return customers! \\
\rowcolor{mygray}
4 & 4.5 & It fits my mom very well! \\
2 & 3.5 & Excellent quality and reasonable price! \\
\rowcolor{mygray}
5 & 4.5 & I can't really recommend it to you! \\
5 & 5 & Cheap and comfortable! \\
\rowcolor{mygray}
5 & 5 & Good! I will buy another one! \\
\hline
\end{tabular}
\end{table}

\noindent It can be seen from Table \ref{table1} that differences do exist between initial rating and improved rating. If the additional comments are negative, the improved ratings will be lower than the initial one. Otherwise, the rating will be greater. In conclusion, the changes of the ratings are consistent with the additional comments and the improved ratings method has been proved.
\subsection{Experiment of recommendation accuracy}
\subsubsection{Data}
\begin{itemize}[leftmargin=0 pt]
    \item Dataset\\
We use MovieLens dataset to evaluate our algorithm (\url{https://movielens.org/}). It is a public available dataset that consists of 100000 ratings from 943 users on 1682 movies. Each user has rated at least 20 objects and made a rating on a scale of 1 to 5.
    \item Data Processing\\
Users’ basic information contained in MovieLens is shown as follows

\begin{table}[H]
\footnotesize
\centering
\caption{The evaluation of improving ratings}\label{table2}
\begin{tabular}{cccc}
\hline
User ID & Age & Gender & Occupation \\
\hline
\rowcolor{mygray}
... & ... & ... & ... \\
\hline
\end{tabular}
\end{table}

For the sake of calculating conveniently, we turn information into digital form. Specifically, the age section is divided into seven groups: below 18, 18-25, 26-35, 36-45, 46-50, 51-56 and above 56, respectively with the representative integers from 1 to 7. Similarly, we represent 21 occupations with the integers 1 to 21, and genders with 0 and 1.Thus users’ basic information can be vectorized.

\begin{table}[H]
\footnotesize
\centering
\caption{Process rule of u.user}\label{table3}
\begin{tabular}{cccccc}
\hline
Age & Quantized Value & Gender & Quantized Value & Occupation & Quantized Value \\
\hline
\rowcolor{mygray}
$<18$ & 1 & male & 1 & administrator & 1 \\
$\cdots$ & $\cdots$ & female & 2 & $\cdots$ & $\cdots$ \\
\rowcolor{mygray}
$>56$ & 7 & $\cdots$ & $\cdots$ & doctor & 21 \\
\hline
\end{tabular}
\end{table}

According to the Bayesian model in WSBM, we figure out the optimal result: 4 block structures. After community detection, we get four communities the number of which is 273, 228, 236 and 206 respectively.
\end{itemize}
\subsubsection{The accuracy of recommendation}
At present, there are two commonly used types of evaluation metrics to evaluate the quality of recommendation system, prediction accuracy and classification accuracy. Prediction accuracy such as MAE and RMSE is to measure the compact degree between the predicted rating and actual rating. Classification accuracy such as Precision, recall and F-measure is to measure how accurately it can predict whether users like or dislike \cite{bobadilla2013recommender}. Without measure the accuracy of prediction directly, any deviation is allowed as long as it has no effect on classification.

Mean absolute error (MAE), it measures the quality of recommendation by calculating the mean absolute error between the actual rating to predicted rating. We obtained the users’ predicted rating values set $\{p_1,\ldots,p_N\}$ and the actual rating values set $\{q_1,\ldots,q_N\}$. MAE is defined as follows
\begin{align}
MAE=\frac{\sum\limits_{i=1}^N\vert p_i-q_i\vert}{N}.
\end{align}

To define the precision, recall and F-measure, we should firstly classify the items which have not been chosen or scored by users. There are four possibilities for a single item: The recommendations that users actually like are classified as True Positive (TP), and the others are classified as False Positive (FP). (For a five-grade marking system, the score of user-liked items is not less than 3).The items that are not recommended but that users actually like are classified as False Negative (FN), and the others are classified as True Negative (TN).

\begin{table}[H]
\footnotesize
\centering
\caption{Four possibilities for a single item}\label{table4}
\begin{tabular}{lll}
\hline
Preference & Recommended & Not Recommended \\
\hline
\rowcolor{mygray}
Like & True Positive (TP) & False Negative (FN) \\
Dislike & False Positive (FP) & True Negative (TN) \\
\hline
\end{tabular}
\end{table}

Precision is defined as the ratio of the number of items which users like as well as recommended by the system to the total number of the recommendation list.
\begin{align}
Precision=\frac{1}{N}\sum_u\frac{TP}{TP+FP}.
\end{align}

Recall is defined as the ratio of the number of items which users like as well as recommended by the system to the total number of user-liked items. Therefore, the recall is given by
\begin{align}
Recall=\frac{1}{N}\sum_u\frac{TP}{TP+FN}.
\end{align}

Therefore we get the F-measure
\begin{align}
\text{F-measure}=\frac{2\times Precision\times Recall}{Precision+Recall}.
\end{align}
\subsubsection{Performance of the method}
Considering the score prediction and classification accuracy of the algorithm, we use all of the evaluation metrics mentioned to evaluate the performance of our recommendation.

As different divisions of training test have great impact on the algorithm accuracy, so we divided MovieLens into training set and test set according to different ratio and then compare MAE and RMSE of conventional and WSBM-based hybrid recommendation. According to the existing research \cite{del2008evaluation}, precision, recall and F-measure have strong dependence on the length of recommendation. The evaluation metrics will have large variation if the length of recommendation list changes, so we fix the proportion of training set and test set (8:2) to compare precision, recall and F-measure of conventional and WSBM-based hybrid recommendation for different length of list. We define variable $\alpha, \beta, \gamma_1, \gamma_2$ as $0.2, 0.8, 0.4, 0.6$ and and obtain the following result.

\begin{figure}[H]
\centering
\includegraphics[width=0.48\textwidth]{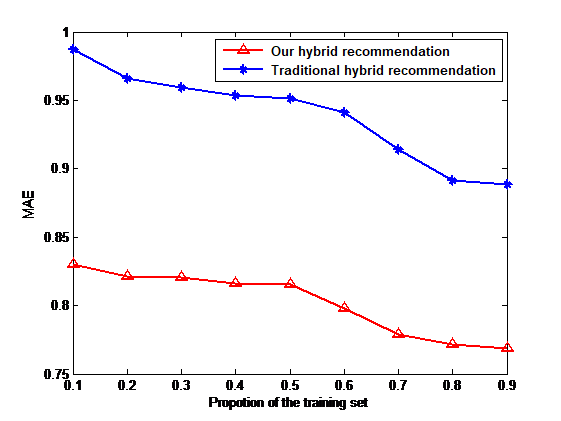}
\caption{The comparison of MAE between our method and traditional method}
\label{fig.mae}
\end{figure}

\begin{figure}[H]
\centering
\includegraphics[width=0.48\textwidth]{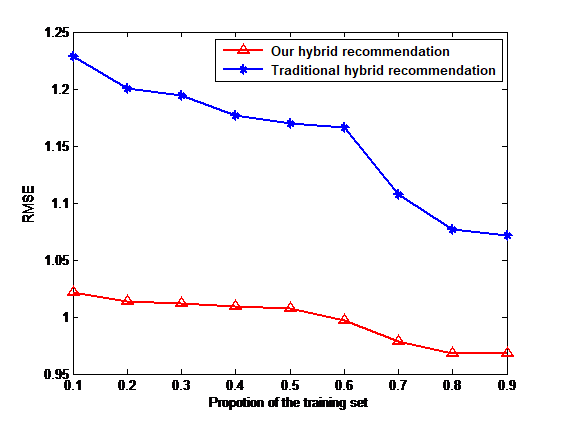}
\caption{The comparison of RMSE between our method and traditional method}
\label{fig.rmse}
\end{figure}

Table \ref{table5} and Table \ref{table6} show the MAE and RMSE of our hybrid recommendation based on WSBM and the traditional hybrid recommendation respectively. The tendency of evaluation metrics change with the different proportion of the training set can be shown in the line charts (Figure \ref{fig.mae} and Figure \ref{fig.rmse}). Define $P^*$ is the proportion of training set and test set.

\begin{table}[H]
\footnotesize
\centering
\caption{MAE of our method and traditional method for MovieLens}\label{table5}
\begin{tabular}{cccccccccc}
\hline
$P^*$ & 0.1 & 0.2 & 0.3 & 0.4 & 0.5 & 0.6 & 0.7 & 0.8 & 0.9 \\
\hline
\rowcolor{mygray}
Ours & 0.8299 & 0.8213 & 0.8202 & 0.8161 & 0.8153 & 0.7977 & 0.7790 & 0.7717 & 0.7683 \\
Traditional & 0.9869 & 0.9961 & 0.9593 & 0.9535 & 0.9515 & 0.9409 & 0.9141 & 0.8918 & 0.8884 \\
\rowcolor{mygray}
Improvement ($\%$) & 15.91 & 14.99 & 14.50 & 14.41 & 14.31 & 15.22 & 14.78 & 13.44 & 13.52 \\
\hline
\end{tabular}
\end{table}

Table \ref{table5} and Figure \ref{fig.mae} show that when the proportion of the training set varies from 0.1 to 0.9, MAE reduce in both methods and reach the lowest points (0.7683 in our hybrid recommendation and 0.8884 in traditional one). So, our hybrid recommendation performs better than the traditional one. For example, our method reduce MAE by $15.91\%$ compared with traditional when $P^*=0.1$.

\begin{table}[H]
\footnotesize
\centering
\caption{RMSE of our method and traditional method for MovieLens}\label{table6}
\begin{tabular}{cccccccccc}
\hline
$P^*$ & 0.1 & 0.2 & 0.3 & 0.4 & 0.5 & 0.6 & 0.7 & 0.8 & 0.9 \\
\hline
\rowcolor{mygray}
Ours & 1.0217 & 1.0134 & 1.0118 & 1.0095 & 1.0071 & 0.9966 & 0.9781 & 0.9707 & 0.9678 \\
Traditional & 1.2288 & 1.2005 & 1.1940 & 1.1769 & 1.1699 & 1.1664 & 1.1072 & 1.0766 & 1.0712 \\
\rowcolor{mygray}
Improvement ($\%$) & 16.85 & 15.59 & 15.26 & 14.22 & 13.92 & 14.56 & 11.66 & 9.84 & 9.65 \\
\hline
\end{tabular}
\end{table}

Table \ref{table6} and Figure \ref{fig.rmse} indicate the increase of the proportion of training set from 0.1 to 0.9 results in decrease of RMSE in both methods. When the proportion of training set reaches 0.9, RMSE fall to 0.9678 in our method and 1.0712 in traditional method. It can be seen from the figure that our method achieves a lower RMSE than traditional method. When $P^*=0.1$, RMSE of our method is 16.85$\%$ lower than that of traditional method. 

To sum up, a larger size of the training set will lead to a better result in recommendation. Compared to the traditional hybrid recommendation, our method achieves better MAE and RMSE which means that selecting the nearest neighbors by WSBM improves the quality of recommendation.

\begin{figure}[H]
\centering
\includegraphics[width=0.48\textwidth]{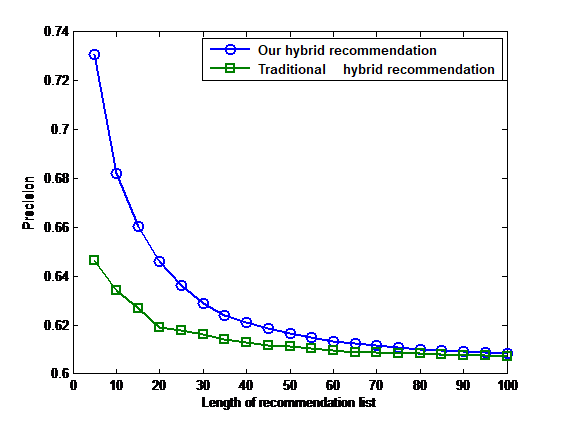}
\caption{The comparison of precision between our method and traditional method}
\label{fig.comparepre}
\end{figure}

\begin{figure}[H]
\centering
\includegraphics[width=0.48\textwidth]{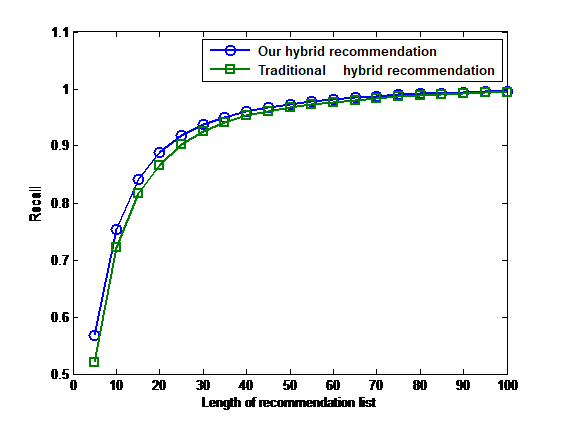}
\caption{The comparison of recall between our method and traditional method}
\label{fig.comparerecall}
\end{figure}

\begin{figure}[H]
\centering
\includegraphics[width=0.48\textwidth]{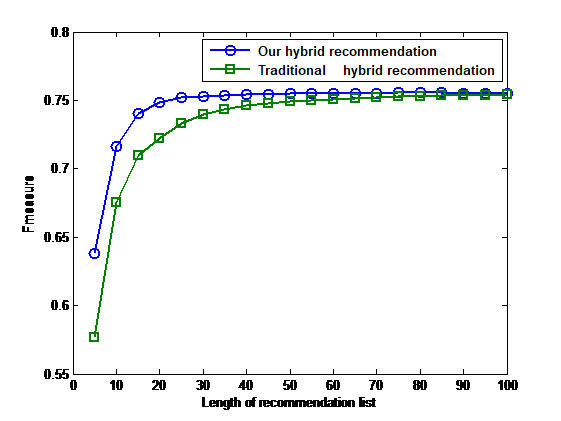}
\caption{The comparison of F-measure between our method and traditional method}
\label{fig.compareF}
\end{figure}

Table \ref{table7}, Table \ref{table8} and Table \ref{table9} show the precision, recall and F-measure of our hybrid recommendation based on WSBM and the traditional hybrid recommendation respectively. We also draw the line charts of them to learn the effects of different length of recommendation list (Figure \ref{fig.comparepre}, Figure \ref{fig.comparerecall} and Figure \ref{fig.compareF}). Define $L^*$ is the length of the recommendation list.

\begin{table}[H]
\footnotesize
\centering
\caption{Precision of our method and traditional method for MovieLens}\label{table7}
\begin{tabular}{ccccccccccc}
\hline
$L^*$ & 5 & 10 & 15 & 20 & 25 & 30 & 35 & 40 & 45 & 50 \\
\hline
\rowcolor{mygray}
Ours & 0.7307 & 0.6823 & 0.6604 & 0.6460 & 0.6361 & 0.6290 & 0.6240 & 0.6210 & 0.6185 & 0.6166 \\
Traditional & 0.6466 & 0.6343 & 0.6270 & 0.6192 & 0.6176 & 0.6160 & 0.6142 & 0.6130 & 0.6116 & 0.6111 \\
\rowcolor{mygray}
Improvement ($\%$) & 13.02 & 7.57 & 5.33 & 4.33 & 3.00 & 2.12 & 1.58 & 1.30 & 1.12 & 0.91 \\
\hline
$L^*$ & 55 & 60 & 65 & 70 & 75 & 80 & 85 & 90 & 95 & 100 \\
\hline
\rowcolor{mygray}
Ours & 0.6148 & 0.6135 & 0.6126 & 0.6118 & 0.6110 & 0.6102 & 0.6097 & 0.6092 & 0.6088 & 0.6085 \\
Traditional & 0.6103 & 0.6098 & 0.6090 & 0.6089 & 0.6084 & 0.6082 & 0.6080 & 0.6076 & 0.6075 & 0.6074 \\
\rowcolor{mygray}
Improvement ($\%$) & 0.74 & 0.61 & 0.59 & 0.47 & 0.29 & 0.26 & 0.22 & 0.19 & 0.43 & 0.33 \\
\hline
\end{tabular}
\end{table}

It can be seen from Table \ref{table7} and Figure \ref{fig.comparepre} that as $L^*$ increases, precision of the two methods decrease from 0.7307 in our method and 0.6466 in traditional method and finally converge to a fixed value. The precision of our method is always higher than that of traditional method, indicating a better accuracy of our method. When $L^*=5$, our method improves the precision by 13.02$\%$ compared with the traditional method.

\begin{table}[H]
\footnotesize
\centering
\caption{Recall of our method and traditional method for MovieLens}\label{table8}
\begin{tabular}{ccccccccccc}
\hline
$L^*$ & 5 & 10 & 15 & 20 & 25 & 30 & 35 & 40 & 45 & 50 \\
\hline
\rowcolor{mygray}
Ours & 0.5668 & 0.7531 & 0.8418 & 0.8885 & 0.9187 & 0.9375 & 0.9506 & 0.9603 & 0.9676 & 0.9732 \\
Traditional & 0.5203 & 0.7221 & 0.8169 & 0.8662 & 0.9022 & 0.9255 & 0.9414 & 0.9527 & 0.9607 & 0.9674 \\
\rowcolor{mygray}
Improvement ($\%$) & 8.93 & 4.30 & 3.04 & 2.58 & 1.84 & 1.29 & 0.98 & 0.8 & 0.72 & 0.6 \\
\hline
$L^*$ & 55 & 60 & 65 & 70 & 75 & 80 & 85 & 90 & 95 & 100 \\
\hline
\rowcolor{mygray}
Ours & 0.9776 & 0.9812 & 0.9844 & 0.9870 & 0.9896 & 0.9913 & 0.9928 & 0.9938 & 0.9948 & 0.9956 \\
Traditional & 0.9727 & 0.9770 & 0.9803 & 0.9839 & 0.9865 & 0.9890 & 0.9907 & 0.9919 & 0.9932 & 0.9942 \\
\rowcolor{mygray}
Improvement ($\%$) & 0.51 & 0.43 & 0.42 & 0.32 & 0.32 & 0.24 & 0.21 & 0.19 & 0.16 & 0.14 \\
\hline
\end{tabular}
\end{table}

Table \ref{table8} and Figure \ref{fig.comparerecall} show us that our hybrid recommendation get a better recall compared with the traditional hybrid recommendation. The increase of the length of the recommendation list from 5 to 30 leads to a dramatically raise of recall and finally the recall will converge to 1 in both methods.

\begin{table}[H]
\footnotesize
\centering
\caption{F-measureof our method and traditional method for MovieLens}\label{table9}
\begin{tabular}{ccccccccccc}
\hline
$L^*$ & 5 & 10 & 15 & 20 & 25 & 30 & 35 & 40 & 45 & 50 \\
\hline
\rowcolor{mygray}
Ours & 0.7402 & 0.7481 & 0.7518 & 0.7529 & 0.7534 & 0.7543 & 0.7546 & 0.7549 & 0.7402 & 0.7481 \\
Traditional & 0.7094 & 0.7222 & 0.7333 & 0.7397 & 0.7434 & 0.746 & 0.7474 & 0.749 & 0.7094 & 0.7222 \\
\rowcolor{mygray}
Improvement ($\%$) & 9.68 & 5.67 & 4.16 & 3.46 & 2.46 & 1.75 & 1.32 & 1.10 & 0.95 & 0.78 \\
\hline
$L^*$ & 55 & 60 & 65 & 70 & 75 & 80 & 85 & 90 & 95 & 100 \\
\hline
\rowcolor{mygray}
Ours & 0.7549 & 0.755 & 0.7552 & 0.7554 & 0.7555 & 0.7554 & 0.7555 & 0.7553 & 0.7554 & 0.7554 \\
Traditional & 0.75 & 0.7509 & 0.7513 & 0.7522 & 0.7526 & 0.7532 & 0.7535 & 0.7536 & 0.7539 & 0.7541 \\
\rowcolor{mygray}
Improvement ($\%$) & 0.64 & 0.54 & 0.51 & 0.42 & 0.38 & 0.29 & 0.26 & 0.22 & 0.19 & 0.17 \\
\hline
\end{tabular}
\end{table}

It can be seen from Table \ref{table9} and Figure \ref{fig.compareF} that there are rapid raises when $L^*$ changes from 5 to 25 and the values converge to a number around 0.75. Our method outperforms the traditional method. When $L^*=5$, our method improves the F-measure by 9.68$\%$ compared with the traditional method.

In conclusion, curves of our method are always above those of the traditional method. When $L^*$ is small, obviously differences do exist between two methods which indicates that our method achieve a better classification accuracy when the recommended items are limited, that is, our method can distinguish the items which users like or dislike more accurately.

\section{Conclusions}
In this paper, we have solved the cold-start problem of new users and new items by combining content-based and collaborative filtering algorithm. In the meantime, we apply WSBM algorithm to find communities with similar user preferences. It is more accurate and effective to predict ratings in communities instead of to an entire social network. As the score sparse will decrease the accuracy of recommendation, we improve the similarity by considering the users’ basic information to gain a more accurate recommendation. Experiments show that hybrid recommendation based on WSBM has better prediction and classification accuracy than the traditional method. However, our method mainly aimed at the users’ dominant behavior, such as buying and grading without taking into account the recessive behavior such as browsing and collection. Actually, considering the recessive behavior can achieve more accurate recommendation. In addition, considering the transfer of user’s interests could be the further research.

\section*{Acknowledgements}
This work is supported by Guangdong Province Student’s Platform for Innovation and Entrepreneurship Training Program (Grant No.1055914152), and The Challenge Cup of the Ministry of Education of China (Grant No.15112005).

\small 
\bibliography{Recommendation.bbl}

\section*{Supplementary Material}

Supplementary material (Algorithm codes and matrices files) will appear in the GitHub as soon as possible.

\end{document}